\RequirePackage{fix-cm}
\documentclass[twocolumn,epjc3]{svjour3}  
\smartqed  
\RequirePackage{graphicx}
\journalname{Eur. Phys. J. C}

\usepackage{multirow}
\usepackage{amsmath}
\usepackage{lineno}
\begin{document}

\title{ParticleNet and its application on CEPC Jet Flavor Tagging
}

\author{Yongfeng Zhu\thanksref{addr1}
        \and
        Hao Liang\thanksref{addr2,addr3} 
        \and
        Yuexin Wang\thanksref{addr2,addr3}
        \and
        Huilin Qu\thanksref{addr4}
        \and
        Chen Zhou\thanksref{addr1}
        \and
        Manqi Ruan\thanksref{addr2,addr3}
}

\thankstext{}{e-mail: manqi.ruan@ihep.ac.cn, czhouphy@pku.edu.cn}


\institute{State Key Laboratory of Nuclear Physics and Technology, School of Physics, Peking University, Beijing, 100871, China \label{addr1}
           \and
           Institute of High Energy Physics, Chinese Academy of Sciences, Beijing 100049, China \label{addr2}
           \and
           University of Chinese Academy of Sciences (UCAS), Beijing 100049, China \label{addr3}
           \and
           CERN, EP Department, CH-1211 Geneva 23, Switzerland \label{addr4}
}

\date{Received: date / Accepted: date}

\maketitle

\begin{abstract}
Quarks, except top quark, and gluon would hadronize and fragment into a spray of stable particles, called jet.
Identification of quark flavor is essential for collider experiments in high-energy physics, relying on flavor tagging algorithms.
In this study, using a full simulation of the Circular Electron Positron Collider (CEPC), we investigated the flavor tagging performance of two different algorithms: ParticleNet, based on Graph Neural Network, and LCFIPlus, based on the Gradient Booted Decision Tree.
Compared to LCFIPlus, ParticleNet significantly enhances flavor tagging performance, resulting in a significant improvement in benchmark measurement accuracy, i.e., a 36\% improvement for $\sigma(ZH)\cdot Br(Z\to \nu\bar{\nu}, H\to c\bar{c})$ measurement and a 75\% improvement for $|V_{cb}|$ measurement via W boson decay, respectively, when CEPC operates as a Higgs factory at the center-of-mass energy of 240~GeV and integrated luminosity of 5.6~$ab^{-1}$.
We compared the performance of ParticleNet and LCFIPlus at different vertex detector configurations, 
observing that the inner radius is the most sensitive parameter, followed by material budget and spatial resolution.


\end{abstract}

\section{Introduction}
\label{intro}
\begin{figure}
\centering
    \includegraphics[width=0.9\columnwidth]{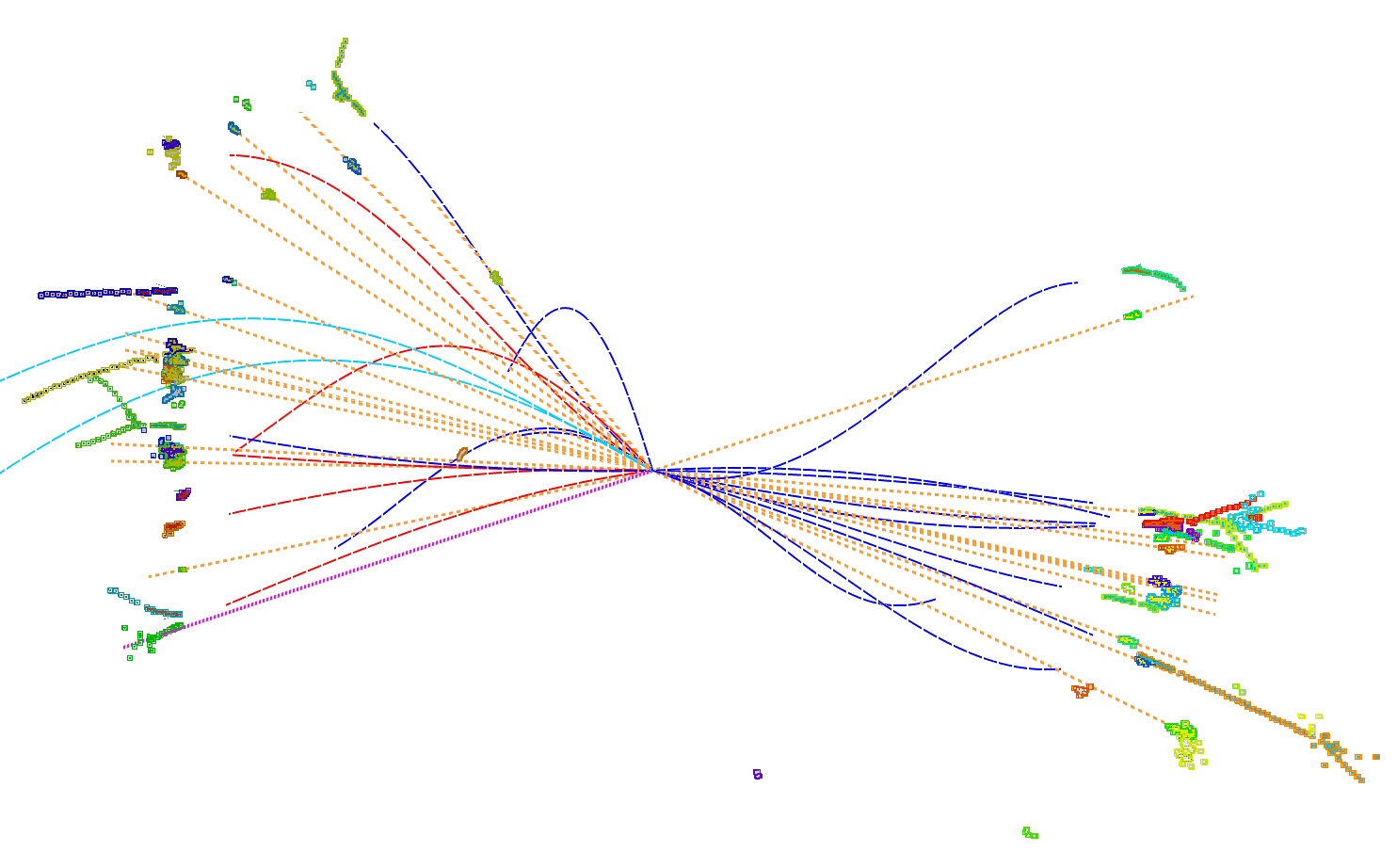}
    \caption{The display of a reconstructed $e^+e^-\to Z\to c\bar{c}$ event with center-of-mass energy of 91.2~GeV. Different particles are depicted with different colors: red for $e^{\pm}$, cyan for $\mu^{\pm}$, blue for $\pi^{\pm}$, orange for photons, and magenta for neutral hadrons. }
    \label{fig:jet}
\end{figure}

A jet refers to a spray of stable particles formed through the hadronization of an energetic quark or gluon. 
The W/Z/Higgs boson and the top quark, the four most massive Standard Model (SM) particles, decay mainly into quarks and hadronize to jets.
Figure~\ref{fig:jet} illustrates a reconstructed $e^+e^-\to Z\to c\bar{c}$ event with center-of-mass energy of 91.2~GeV. 
Efficient identification of the jet flavor could shed light on the properties of those massive particles and is critical for experimental exploration at the high-energy frontier.
Flavor tagging is used to distinguish jets hadronized from different quark flavors or gluon.
To promote the development of future electron-positron Higgs factories, which is regarded as the highest priority for the next collider~\cite{ES}, accurate performance analysis and optimization of both detectors and algorithms are essential.
Jet flavor tagging and relevant benchmark analyses serve as good objectives.

\begin{table*}
\centering
\caption{The operation scheme of the CEPC, including the center-of-mass energy,  the instantaneous luminosity, the total integrated luminosity, and the event yields~\cite{Sun:2023lut}.}
\label{CECPope}
\begin{tabular*}{\textwidth}{@{\extracolsep{\fill}}ccccc@{}}
\hline
Operation mode    & Z factory     & WW     & Higgs factory    & $t\bar{t}$ \\
\hline
$\sqrt{s}$ (GeV) & 91.2   & 160     & 240      & 360 \\
Run time (year)  & 2 & 1 & 10 & 5 \\
Instantaneous luminosity      &   \multirow{2}{*}{191.7}    &   \multirow{2}{*}{26.6}    &   \multirow{2}{*}{8.3}   &  \multirow{2}{*}{0.83} \\
($\rm 10^{34}cm^{-2} s^{-1}$, per IP)  & & & & \\
Integrated luminosity      &  \multirow{2}{*}{100}    &   \multirow{2}{*}{6}    &   \multirow{2}{*}{20}   &  \multirow{2}{*}{1} \\
($\rm ab^{-1}$, 2 IPs) & & & & \\
Event yields   &  3$\times$ 10$^{12}$  & 1$\times$ 10$^8$ & 4$\times$ 10$^6$ & 5 $\times$ 10$^{5}$ \\
\hline
\end{tabular*}
\end{table*}

The Circular Electron Positron Collider (CEPC)~\cite{CEPCStudyGroup:2018ghi} is a large-scale collider facility proposed after the discovery of the Higgs boson in 2012. 
It is designed with a circumference of 100~km with two interaction points.
It can operate at multiple center-of-mass energies, including 240~GeV as a Higgs factory, 160~GeV for the $\rm W^+W^-$ threshold scan, 
 and 91~GeV as a Z factory.
It also can be upgraded to 360~GeV for the $t\bar{t}$ threshold scan. 
Table~\ref{CECPope} summarizes its baseline operating scheme and the corresponding boson yields~\cite{Sun:2023lut}.
In the future, it can be upgraded to a proton-proton collider to directly explore new physics at a center-of-mass energy of about 100~TeV.
The main scientific objective of the CEPC is the precise measurement of the Higgs properties, especially its coupling properties.
Additionally, trillions of $Z\to q\bar{q}$ events can provide an excellent opportunity for studying flavor physics.

Jet flavor tagging performance depends on detector design, particularly the design of the vertex detector, as well as the utilization of reconstruction algorithms. 
In this study, we apply ParticleNet~\cite{Qu:2019gqs} to the CEPC and assess its flavor tagging performance in the measurement of $\sigma(ZH)\cdot Br(Z\to \nu\bar{\nu}, H\to c\bar{c})$ and $|V_{cb}|$ via W decay.
Our results demonstrate that ParticleNet outperforms the baseline jet flavor tagging algorithm, LCFIPlus~\cite{LCFIPlus}, by achieving a 36\% and 75\% improvement in the relative statistical accuracy of $\sigma(ZH)\cdot Br(Z\to \nu\bar{\nu}, H\to c\bar{c})$ and $|V_{cb}|$ measurement via W boson decay at the center-of-mass energy of 240~GeV and integrated luminosity of 5.6~$ab^{-1}$.
We also observe that both ParticleNet and LCFIPlus perform better in the barrel region when compared to the endcap region.
By analyzing the dependence of flavor tagging performance on vertex detector configurations, we observe that the most sensitive vertex detector parameter is the inner radius, followed by the material budget and spatial resolution.
This result is consistent with previous studies conducted using LCFIPlus.

This article is organized as follows. 
Section~\ref{CEPC} introduces the CEPC detector, software, and the samples used in this analysis.
Section~\ref{FT} describes the jet flavor tagging algorithms (LCFIPlus and ParticleNet) and the method used to evaluate their performance.
Section~\ref{perform} quantifies the dependence of flavor tagging performance on vertex detector configuration and compares the performance of ParticleNet and LCFIPlus.
Finally, Section~\ref{sec:Con} provides a brief conclusion.

\section{CEPC Detector, software, and samples}
\label{CEPC}

\begin{figure}
\centering
    \includegraphics[width=0.95\columnwidth]{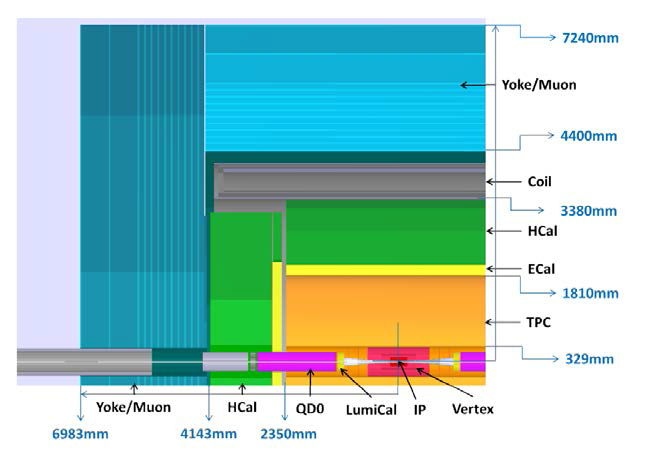}
    \caption{The CEPC baseline detector. From inner to outer, the detector is composed of a silicon pixel vertex detector, a silicon inner tracker, a TPC, a silicon external tracker, an ECAL, an HCAL, a solenoid of 3 Tesla, and a return yoke embedded with a muon detector. Five pairs of silicon tracking disks are installed in the forward regions to enlarge the tracking acceptance.~\cite{Sun:2023lut}}
    \label{fig:Det}
\end{figure}

\begin{figure}
\centering
    \includegraphics[width=0.7\columnwidth]{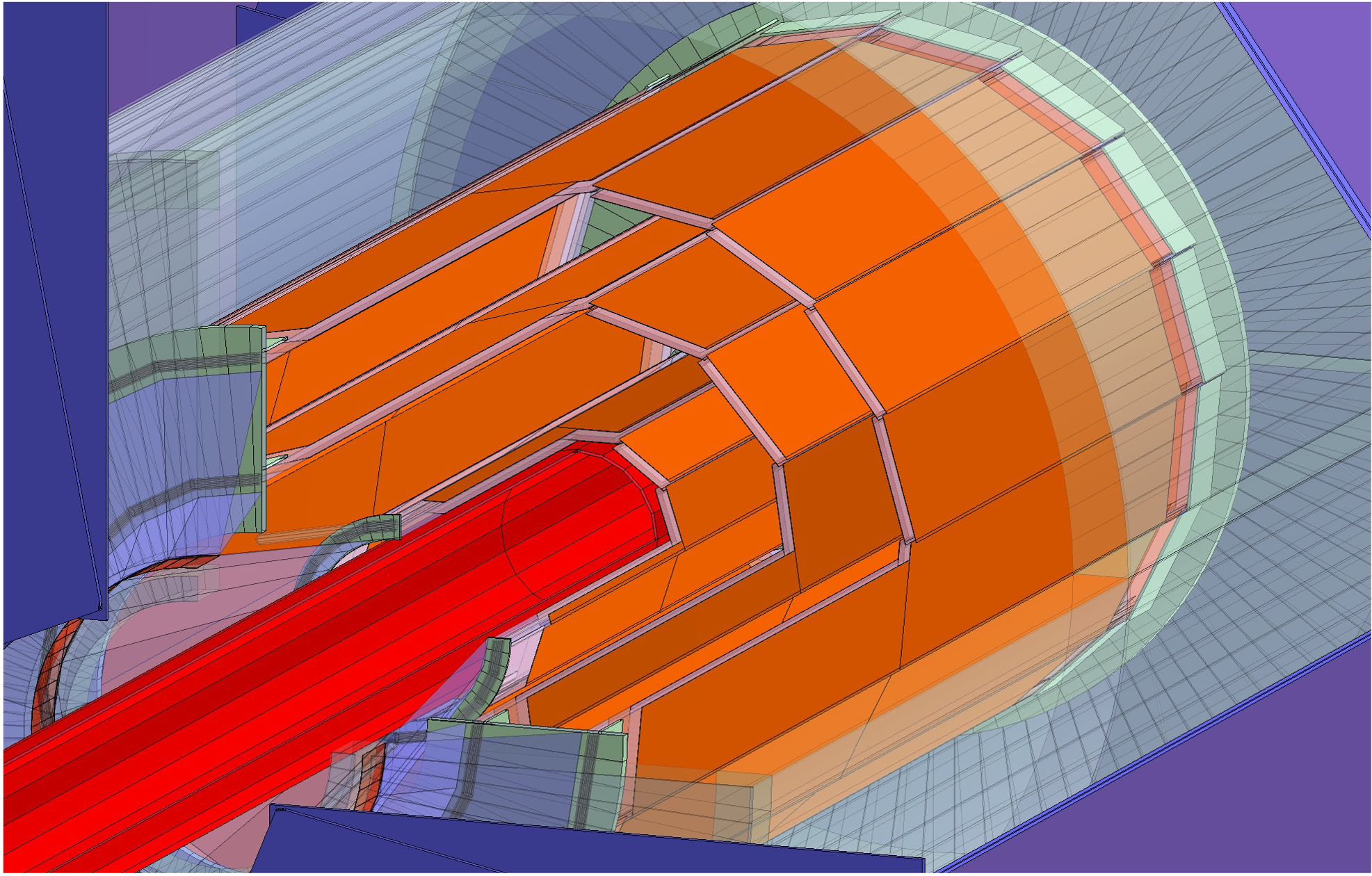}
    \caption{Schematic view of vertex detector. Two layers of silicon pixel sensors are mounted on both sides of each of the three ladders to provide six space points. The beam pipe is surrounded by the vertex detector.~\cite{VTX} }
    \label{fig:vertex}
\end{figure}

At present, two interaction points of CEPC are designed with the same baseline detector, which is designed according to the Particle Flow Algorithm (PFA) principle and emphasizes reconstructing visible final state particles in the most-suited detector subsystems.
The structure of the CECP detector is shown in Fig.~\ref{fig:Det}.
From inner to outer, the baseline detector is composed of a silicon pixel vertex detector, a silicon inner tracker, a Time Projection Chamber (TPC) surrounded by a silicon external tracker, a silicon-tungsten sampling Electromagnetic Calorimeter (ECAL), a steel-glass Resistive Plate Chambers sampling Hadronic Calorimeter (HCAL), a 3 Tesla superconducting solenoid, and a flux return yoke embedded with a muon detector.
For flavor tagging, the vertex detector is critical.
At the CEPC, the vertex detector is designed with six concentric cylindrical layers of square silicon pixel sensors.
The mechanical structure of the vertex detector consists of ladders, with each ladder supporting sensors on both sides. 
The detailed structure of the vertex detector is depicted in Fig.~\ref{fig:vertex}, and its specific parameters are listed in Table~\ref{vtx}.

\begin{table}
\caption{ The baseline design parameters of the CEPC vertex system.~\cite{VTX}}
\label{vtx}
\begin{tabular}{cccc}
\hline
                    &   radius       & spatial resolution      & material budget    \\
                    &   (mm)      & ($\mu m$) & \\
\hline
Layer 1             &   16          & 2.8                              & 0.15\%/X$_0$    \\
Layer 2             &  18           & 6                              & 0.15\%/X$_0$     \\
Layer 3             & 37            & 4                               & 0.15\%/X$_0$   \\
Layer 4             & 39            & 4                               & 0.15\%/X$_0$   \\
Layer 5             &  58           &  4                            & 0.15\%/X$_0$          \\
Layer 6             &  60           &  4                            & 0.15\%/X$_0$          \\
\hline
\end{tabular}
\end{table}

\begin{figure}
\centering
    \includegraphics[width=0.95\columnwidth]{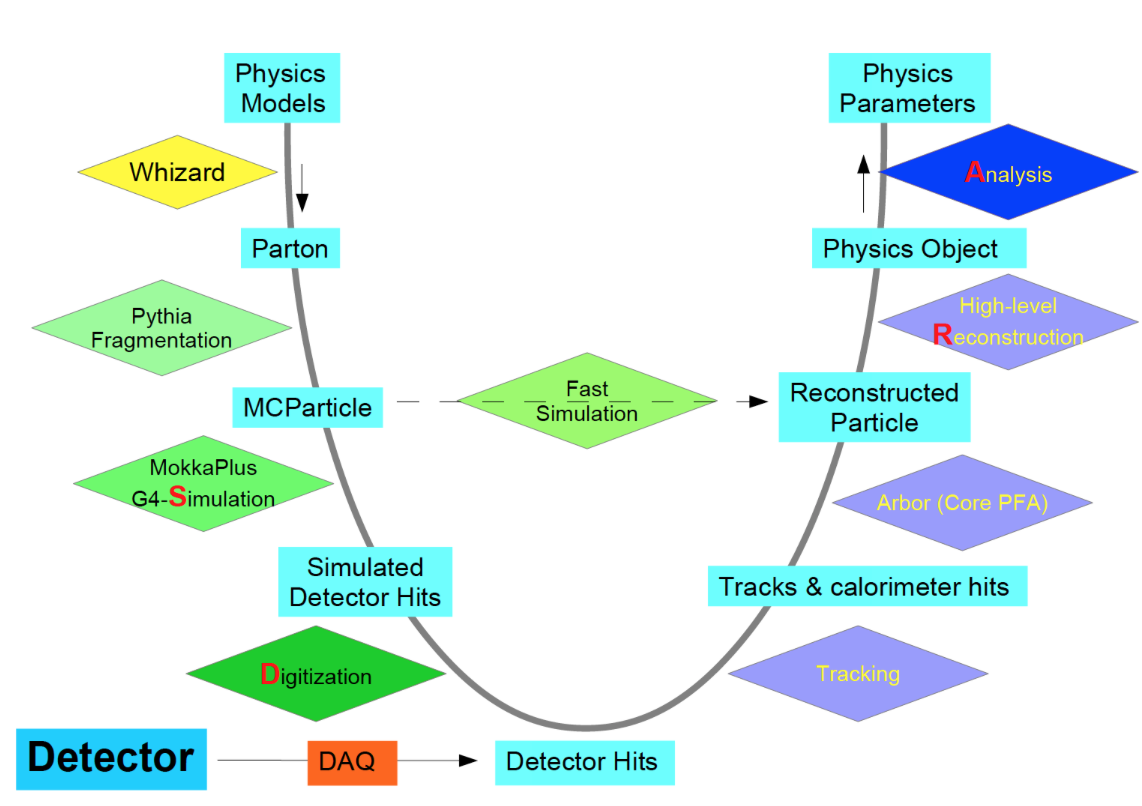}
    \caption{The information flow of the CEPC software chain.~\cite{zhuEPJC} }
    \label{fig:software}
\end{figure}

A baseline reconstruction software chain has been developed to quantify the scientific merit and guide the detector optimization of CEPC, see Fig.~\ref{fig:software}.
The data flow of the CEPC baseline software starts from the event generators of Whizard~\cite{whizard} and Pythia~\cite{pythia}.
The detector geometry is implemented into MokkaPlus~\cite{mokka}, a GEANT4-based full simulation module.
MokkaPlus calculates the energy deposition in the detector-sensitive volumes and creates simulated hits.
For each sub-detector, the digitization module converts the simulated hits into digitized hits by convolution of the corresponding sub-detector responses.
The reconstruction modules include the tracking, the Particle Flow, and the high-level reconstruction algorithms.
The digitized tracker hits are reconstructed into tracks via the tracking algorithms~\cite{track}.
The Particle Flow algorithm, Arbor~\cite{arbor}, reads the reconstructed tracks and the calorimeter hits to build reconstructed particles.
High-level reconstruction algorithms reconstruct composite physics objects such as converted photons, jets, taus, and so on, and identify the flavor of the jets.

In this paper, we utilized hadronic events at Z-pole operation,
including 1 million $Z\to b\bar{b}$ events, 1 million $Z\to c\bar{c}$ events, and 0.33 million each of $Z\to u\bar{u}/d\bar{d}/s\bar{s}$ events.
For ParticleNet, we divided the samples into three distinct sets: the training set for training the model, the validation set used to validate whether the model is overfitting or underfitting, and the testing set used to give flavor tagging results. 
The ratios of samples in these sets were set at 60\%, 20\%, and 20\%, respectively. 
For LCFIPlus, we use all samples to do the test since we have already trained the model.

\section{Flavor tagging algorithms and their performance}
\label{FT}

In this section, we introduce LCFIPlus and ParticleNet and compare their performance based on the CEPC detector and software.
Both algorithms read the information of reconstructed jet candidates and calculate the jet likeness to b, c, and light categories.

The LCFIPlus package, a framework for jet analysis in linear collider studies, was originally developed by the International Linear Collider (ILC)~\cite{ILC}, and has since been widely used at the Compact Linear $e^+e^-$ Collider (CLIC)~\cite{Robson:2018enq}, the Future Circular Collider $e^+e^-$ (FCC-ee)~\cite{FCCCDR}, and CEPC.
The LCFIPlus package consists of vertex finding, jet clustering, vertex refinement, and flavor tagging. 
To perform flavor tagging, the jets are classified into four categories based on the number of reconstructed vertices and isolated leptons in the jet.  
A set of variables is then extracted for each category, which includes the number of tracks in each vertex, the vertex mass, the distance between the secondary vertex and the primary vertex, the vertex decay length, the track transverse momentum, and more.
Further details can be found in~\cite{LCFIPlus}. 
In each category, two types of flavor tagging algorithms are trained using the Gradient Boosted Decision Tree (GBDT) method, one for the b-tagging algorithm and the other for the c-tagging algorithm.

\begin{figure}
\includegraphics[width=.4\columnwidth]{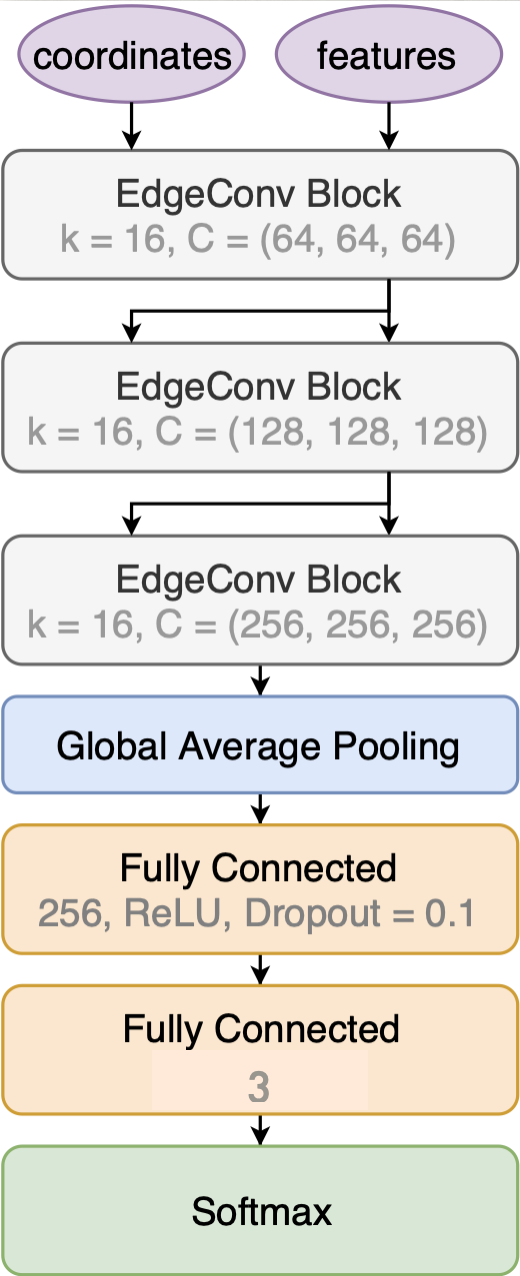}
\qquad
\includegraphics[width=.5\columnwidth]{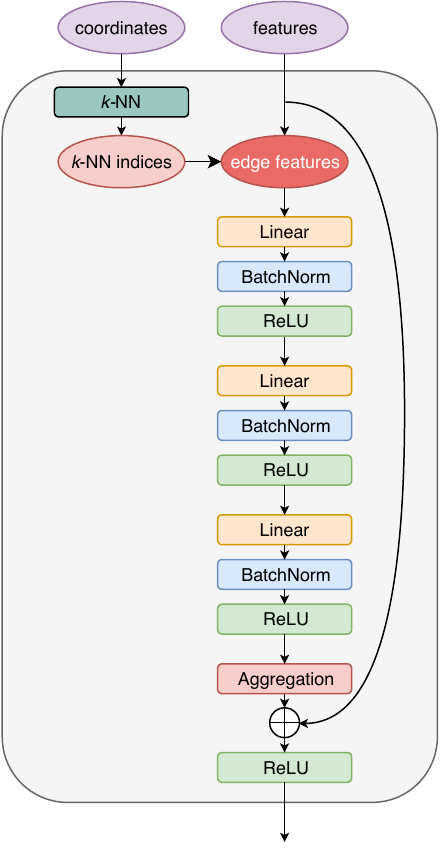}
\caption{The architecture of the ParticleNet (left) and the structure of the EdgeConv block (right).~\cite{Qu:2019gqs} }
\label{PN}
\end{figure}

\begin{table*}
\centering
\caption{ The input variables used in ParticleNet for jet flavor tagging at the CEPC.}
\label{tab:features}
\begin{tabular*}{\textwidth}{@{\extracolsep{\fill}}cc@{}}
\hline
Variable                    &   Definition              \\
\hline
$\Delta \eta$  &   difference in pseudorapidity between the particle and the jet axis     \\
$\Delta \phi$   &   difference in azimuthal angle between the particle and the jet axis        \\
\hline
$\rm log P_t$                   &   logarithm of the particle's $P_t$      \\
$\rm log E$                     &   logarithm of the particle's energy      \\
$\rm log\frac{P_t}{P_t(jet)}$   &   logarithm of the particle's $P_t$ relative to the jet $P_t$       \\
$\rm log\frac{E}{E(jet)}$       &   logarithm of the particle's energy relative to the jet energy       \\
$\Delta R$ &   angular separation between the particle  and the jet axis ($\sqrt{(\Delta \eta)^2 + (\Delta \phi)^2}$)       \\
$d_0$                          &   transverse impact parameter of the track\\
$d_0$err                       &   uncertainty associated with the measurement of the $d_0$\\
$z_0$                          &   longitudinal impact parameter of the track\\
$z_0$err                       &   uncertainty associated with the measurement of the $z_0$\\
charge                      &   electric charge of the particle     \\
\hline
isElectron                  &   whether the particle is an electron      \\
isMuon                      &   whether the particle is a muon       \\
isChargedKaon             &   whether the particle is a charged Kaon      \\
isChargedPion             &   whether the particle is a charged Pion      \\
isProton             &   whether the particle is a proton      \\
isNeutralHadron             &   whether the particle is a neutral hadron       \\
isPhoton                    &   whether the particle is a photon       \\
\hline
\end{tabular*}
\end{table*}

The ParticleNet based on Graph Neural Network (GNN)~\cite{zhou2020graph} was published at the beginning of 2019.
The architecture of ParticleNet is shown in the left plot of Fig.~\ref{PN}.
It consists of three EdgeConv blocks, one channel-wise global average pooling block, and two fully connected blocks followed by a softmax function to output the b/c/light-likeness for each jet.
The core concept of ParticleNet is the EdgeConv operation, which is realized by applying feature aggregation for each particle and its k nearest particles in the jet.
The specific process of each EdgeConv block is illustrated in the right plot of Fig.~\ref{PN}. 
It starts by finding the $k$-nearest neighbors for each particle within the jet. 
The edge between each particle and its $k$-nearest neighbors is determined using the input features of each particle. 
In the first EdgeConv block, the spatial coordinates ($\Delta \eta, \Delta \phi$) of the particles in the pseudorapidity-azimuth space are used to compute the edge of each pair of particles, while the subsequent EdgeConv blocks use the learned feature vectors as coordinates.
The input features for our task, listed in Table~\ref{tab:features}, include the kinematic variables constructed with the 4-momentum of each particle, the PID information, the charge, and impact parameters.
The distance between the interaction point and the path of a track is defined as the impact parameter, where the distance along the beam is called $z_0$ and perpendicular to the beam is called $d_0$.

\begin{figure*}
\centering
\includegraphics[width=.4\textwidth]{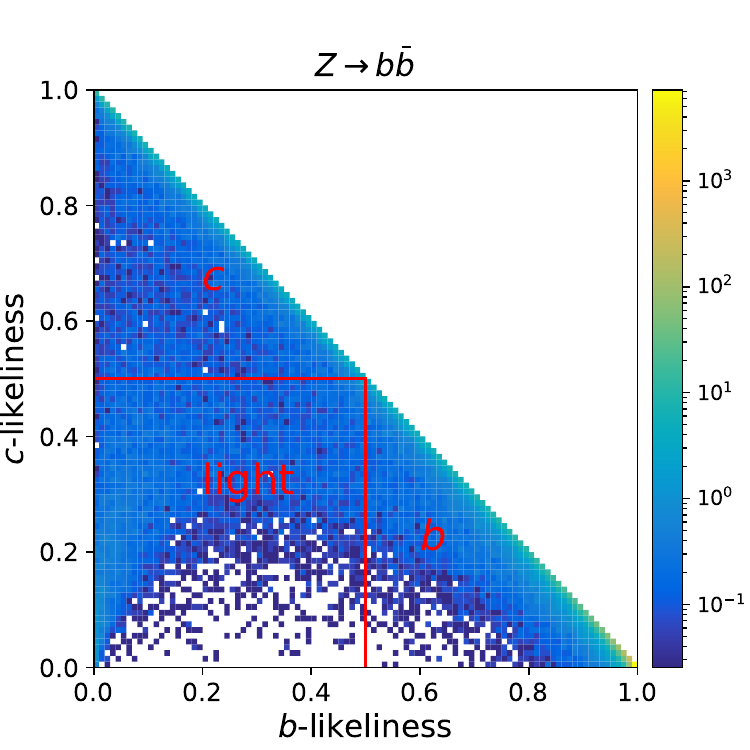}
\\
\includegraphics[width=.4\textwidth]{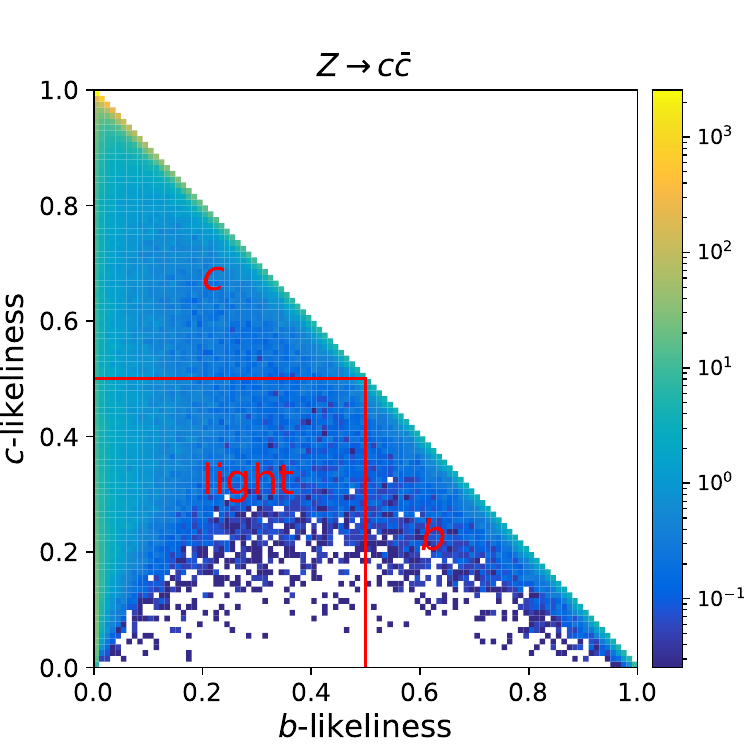}
\qquad
\includegraphics[width=.4\textwidth]{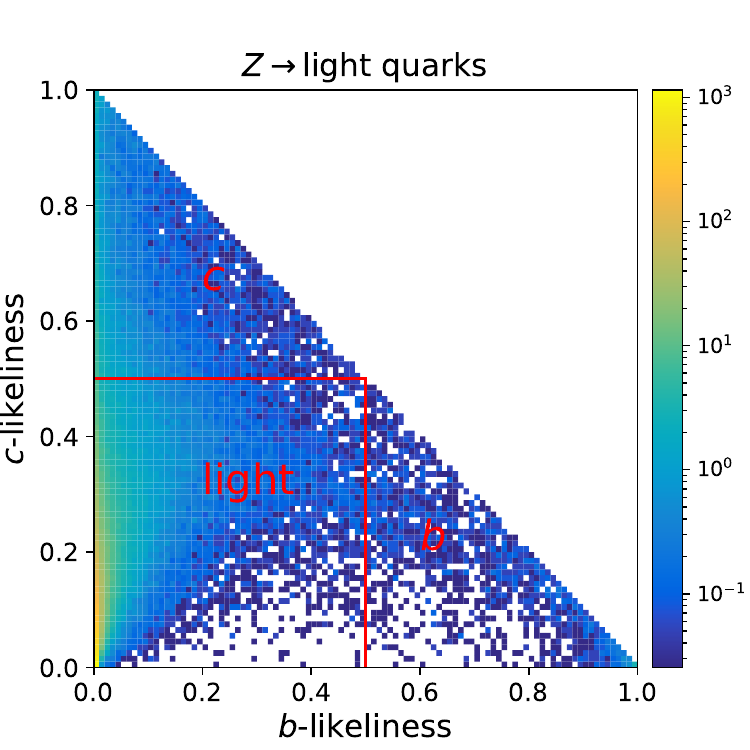}
\caption{The distribution of b/c-likeness for samples of $Z\to b\bar{b}/c\bar{c}/$light quarks. The parallel lines divide the space spanned by the b/c-likeness into three regions. }
\label{FTPerform}
\end{figure*}

\begin{figure}
\centering
\includegraphics[width=.9\columnwidth]{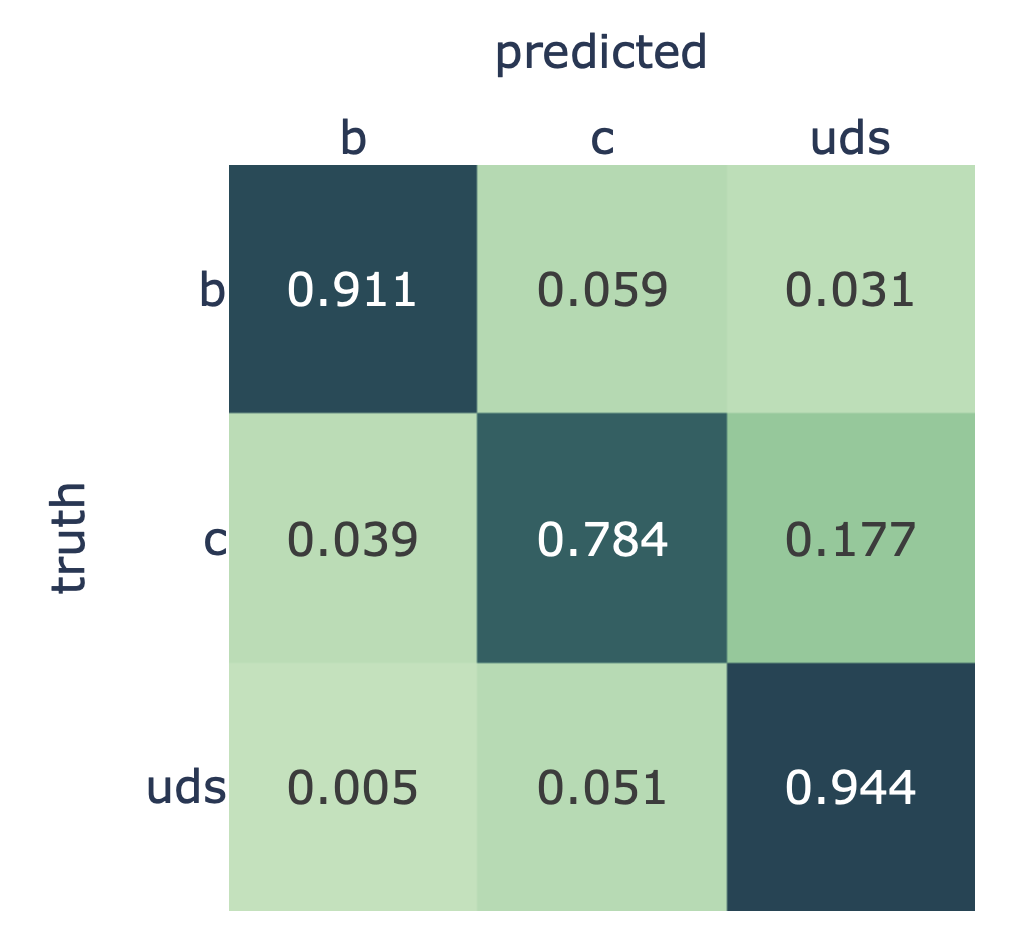}
\\
\includegraphics[width=.9\columnwidth]{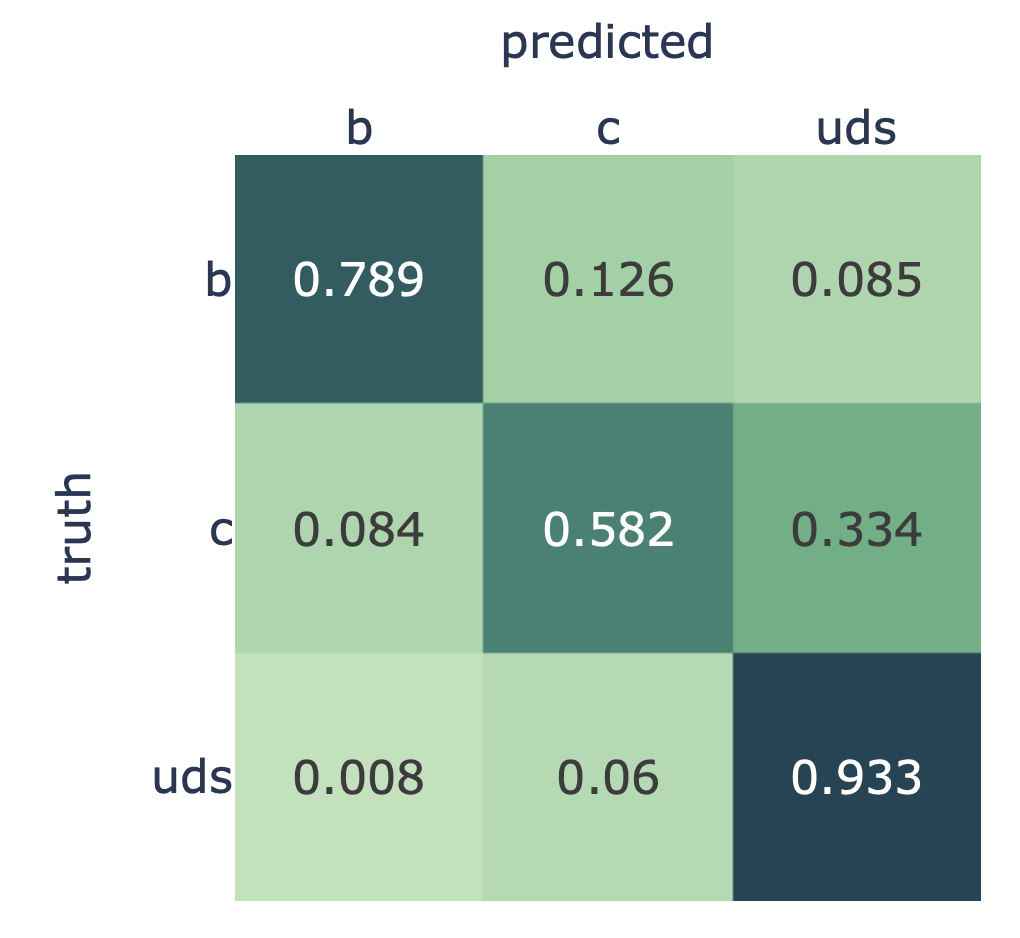}
\caption{The migration matrix of flavor tagging performance of ParticleNet (top) and LCFIPlus (bottom) at the CEPC.}
\label{PerformPNLCFI}
\end{figure}

Both flavor tagging algorithms assign three values to each jet: b-likeness, c-likeness, and light-likeness, with the constraint that their sum equals unity. 
The scatter plots in Fig.~\ref{FTPerform} show the distribution of b-likeness versus c-likeness for samples of $Z\to b\bar{b}/c\bar{c}$/light quarks with ParticleNet. 
In these plots, b-jets tend to concentrate in the region of larger b-likeness, c-jets in the region of larger c-likeness, and light-jets in the region of smaller b/c-likeness.
The phase space spanned by the b/c-likeness is divided into three different regions corresponding to identified b, c, and light quarks.
We then obtain the ratios of b-jets identified as b-jets, b-jets identified as c-jets, and so on.
These ratios can be represented with a migration matrix, as shown in Fig.~\ref{PerformPNLCFI}.
The working point (phase space separation) can be optimized according to the specific analysis requirements.
For general cases, we adopt the method using two orthogonal lines passing through the point (0.5, 0.5), as depicted by the two red lines in figure~\ref{FTPerform}.

\section{Performance analyses}
\label{perform}

The performance of ParticleNet and LCFIPlus is evaluated by the following three criteria.
The first one is the migration matrix since the perfect flavor tagging performance corresponds to the identity matrix.
The second one is the physics performance, the better flavor tagging algorithm would induce better physics results.
The last one is the vertex detector optimization since it is relevant to the resolution of transverse momentum and impact parameters, and further quantifies the reconstructed performance of the detector.

\subsection{Performance comparison and impact on benchmarks of $\sigma(ZH)\cdot Br(Z\to \nu\bar{\nu}, H\to c\bar{c})$ and $|V_{cb}|$}
\label{4.1}

Figure~\ref{PerformPNLCFI} displays the migration matrices obtained using LCFIPlus and ParticleNet, respectively. 
Comparing the performance of LCFIPlus, ParticleNet achieves a significant improvement in b/c-tagging efficiency, with an enhancement of 15\% for b jets and 32\% for c jets.
The trace of the matrix abbreviated as $\rm Tr_{mig}$ is 3.0 for perfect jet flavor tagging performance and it increases from 2.30 to 2.64 with the utilization of ParticleNet.
Both LCFIPlus and ParticleNet face a more challenging task in c-tagging, as its properties lie between those of b and light jets.

\begin{figure}
\includegraphics[width=.9\columnwidth]{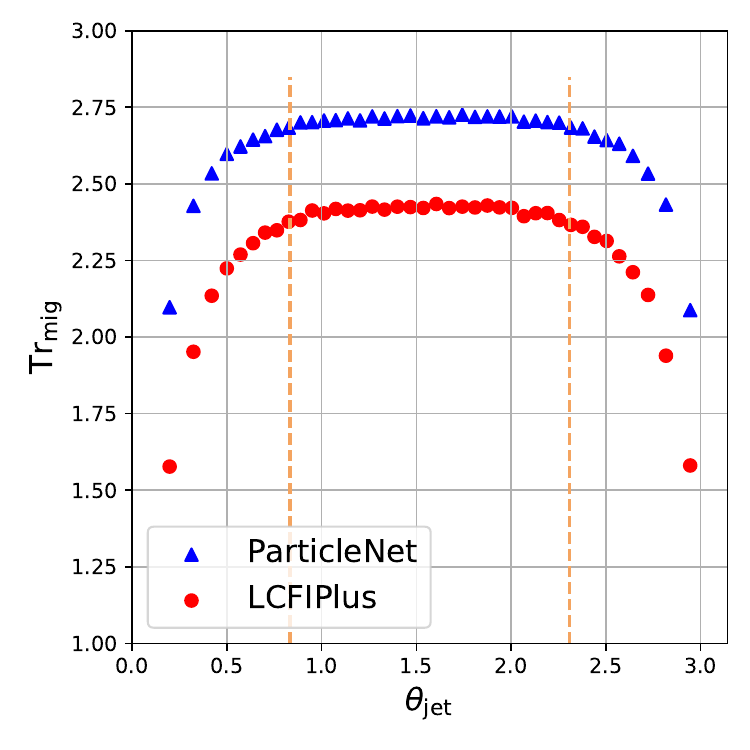}
\\
\includegraphics[width=.9\columnwidth]{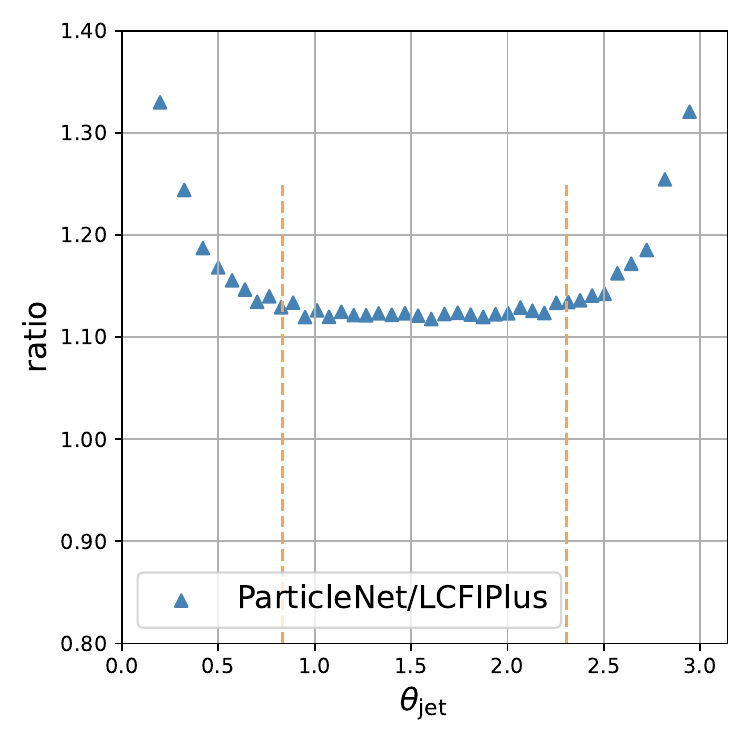}
\caption{The top plot shows the correlation between jet flavor tagging performance, quantified using the trace of the flavor tagging performance matrix, and the jet polar angle. The bottom plot illustrates the performance improvement of ParticleNet relative to LCFIPlus at different jet polar angles. Two vertical lines mark the boundary between the barrel and endcap regions.}
\label{theta}
\end{figure}

In the top plot of Fig.~\ref{theta}, we present the correlation between jet flavor tagging performance, described by $\rm Tr_{mig}$, and jet polar angle, which is defined as the angle with respect to the beam line and represented by the angle $\theta_{jet}$. 
Both LCFIPlus and ParticleNet exhibit better performance in the barrel region compared to the endcap region, due to the relatively lower resolution of transverse momentum ($P_t$) and impact parameters ($d_0$ and $z_0$) in the endcap region.
The value of ParticleNet performance divided by LCFIPlus performance can be used to describe the performance improvement of ParticleNet relative to LCFIPlus.
The bottom plot of Fig.~\ref{theta} shows the correlation between those values and the jet polar angle.
Compared to LCFIPlus, ParticleNet can improve the trace of the migration matrix by more than 10\% in the barrel region and more than 30\% in the endcap regions.

\begin{figure}
\centering
\includegraphics[width=.45\textwidth]{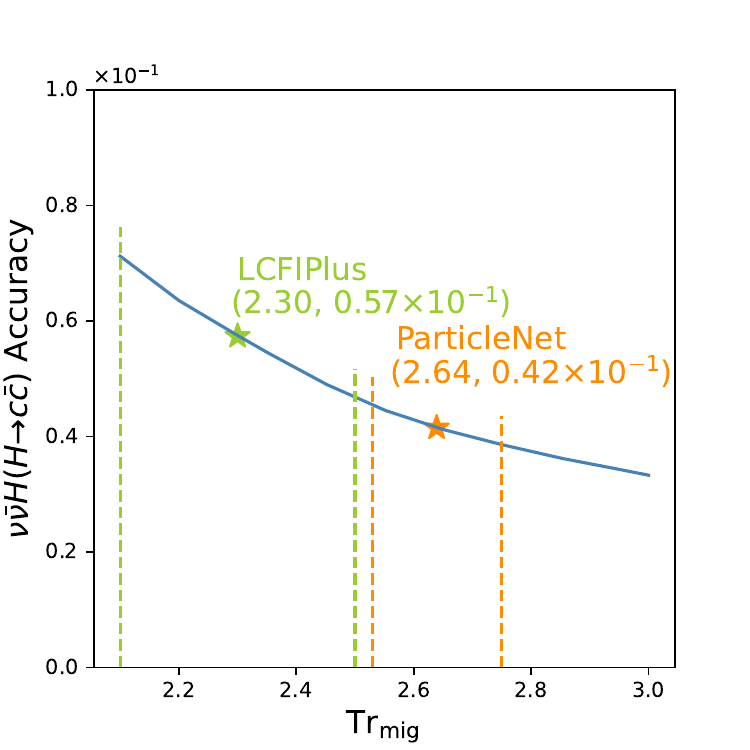}
\\
\includegraphics[width=.45\textwidth]{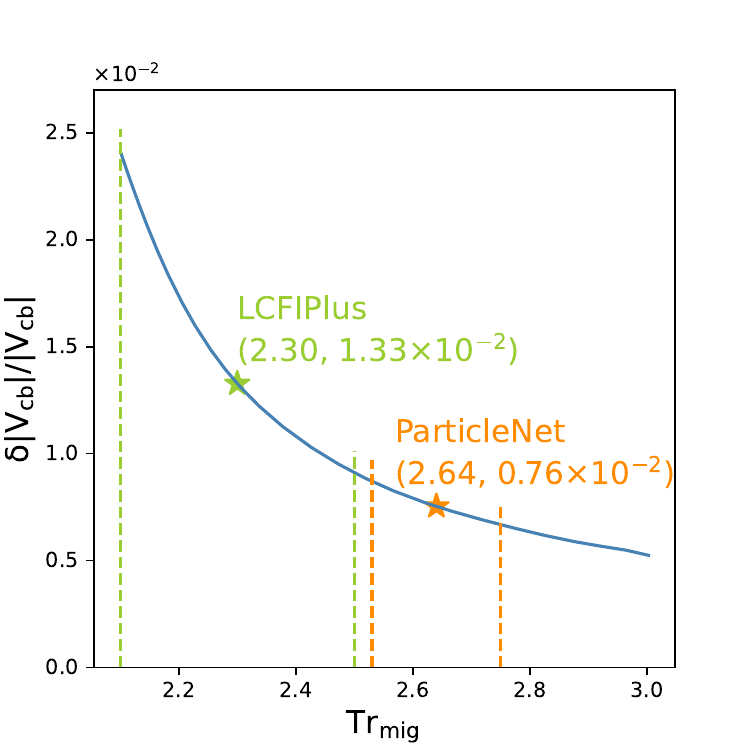}
\caption{The dependence of relative statistical uncertainties for measurement of $\sigma(ZH)\cdot Br(Z\to \nu\bar{\nu}, H\to c\bar{c})$ (top) and $|V_{cb}|$ (bottom) on flavor tagging performance, which is represented with the trace of flavor tagging performance matrix. The larger green/orange marker corresponds to the result obtained by LCFIPlus/ParticleNet. When the vertex detector parameters, including the inner radius, material budget, and spatial resolution, are changed by a factor of 0.5/2 from the baseline design (the geometry used in this simulation), the $\rm Tr_{mig}$ value changes accordingly. It shifts from 2.64 to 2.75/2.53 for ParticleNet and from 2.30 to 2.10/2.50 for LCFIPlus, as indicated by the four vertical lines.}
\label{anaPerform}
\end{figure}

The performance of both flavor tagging algorithms is compared in benchmark analyses.
The first analysis we look into is the signal strength measurement of $\sigma(ZH)\cdot Br(Z\to \nu\bar{\nu}, H\to c\bar{c})$.
In the paper~\cite{HiggsBCG}, the authors demonstrate a correlation between the trace of the migration matrix and the accuracy of the signal strength of $\sigma(ZH)\cdot Br(Z\to \nu\bar{\nu}, H\to c\bar{c})$ when CEPC operates as a Higgs factory at the center-of-mass energy of 240~GeV and integrated luminosity of 5.6~$ab^{-1}$, as depicted in the top plot of Fig.~\ref{anaPerform}.
Using LCFIPlus, the trace is 2.30, corresponding to an accuracy of 0.057, indicated by the green star.
ParticleNet enhances the trace to 2.64, aligning with an accuracy of 0.042, represented by the orange star.
The second analysis is the signal strength measurement of $|V_{cb}|$, the magnitude of $V_{cb}$, which is one of the Cabibbo-Kobayashi-Maskawa (CKM) matrix elements and governs the transition between charm and bottom quarks.
Accurate measurement of $|V_{cb}|$ plays a pivotal role in the study of weak interactions within the SM.
When CEPC operates as a Higgs factory at the center-of-mass energy of 240~GeV and integrated luminosity of 5.6~$ab^{-1}$, the authors demonstrate that ParticleNet can significantly improve the accuracy of signal strength by 75\% in the measurement of $|V_{cb}|$ through $W^+W^-\to \mu\nu q\bar{q}$~\cite{Vcb}, as depicted in the bottom plot of Fig.~\ref{anaPerform}.

\subsection{Comparison on vertex detector optimization}

Jet flavor tagging performance depends on the detector design, especially the vertex detector.
The vertex detector is mainly characterized by three parameters: material budget, spatial resolution, and inner radius.
The CEPC vertex detector is designed with three concentric cylinders of double-sided layers, and the parameters are listed in Table~\ref{vtx}.

\begin{figure}
\centering
\includegraphics[width=.9\columnwidth]{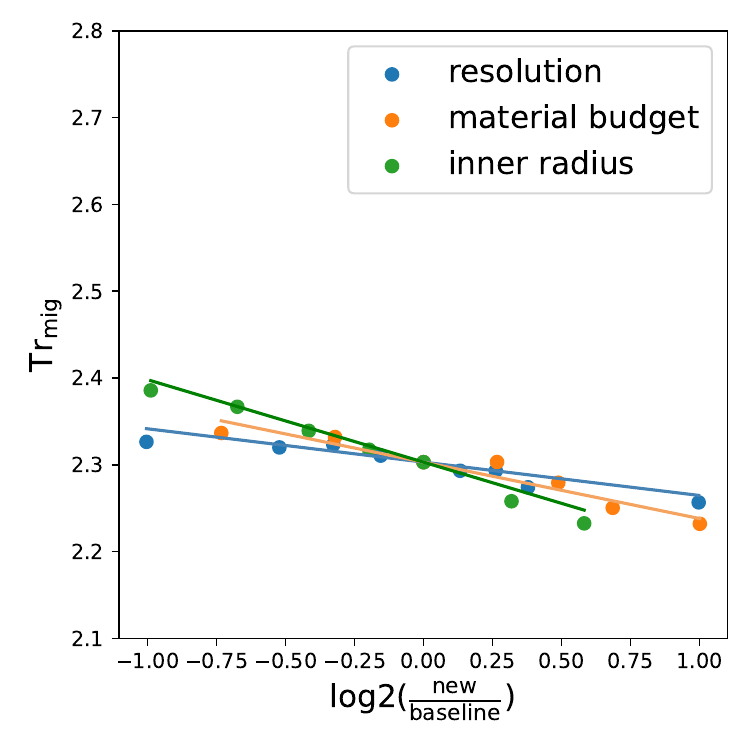}
\\
\includegraphics[width=.9\columnwidth]{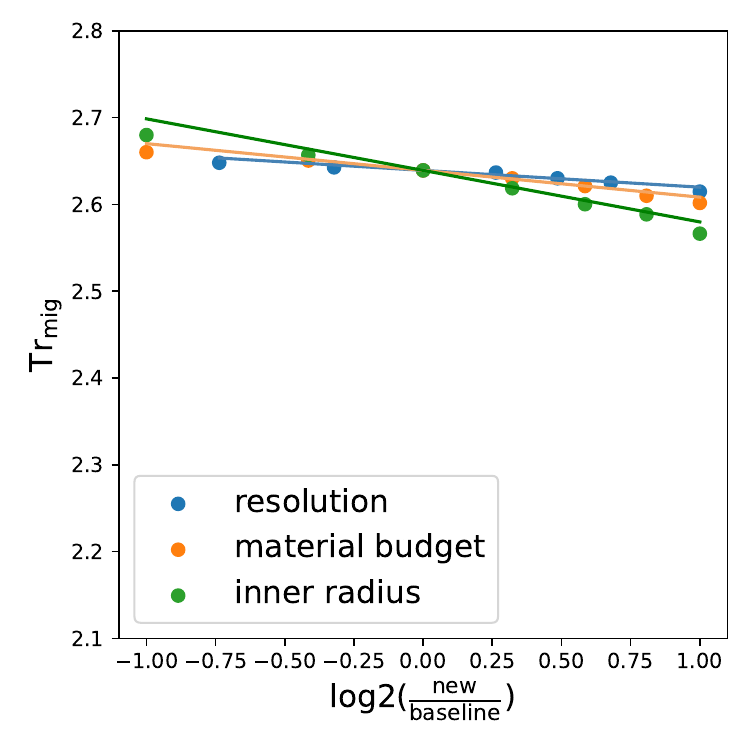}
\caption{The correlation between the trace of a migration matrix and relative scanned parameters for LCFIPlus (top) and 
 ParticleNet (bottom).}
 \label{PNscan}
\end{figure}

In a previous study~\cite{VTX} using the LCFIPlus flavor tagging algorithm, the correlation between c-jet tagging efficiency multiplied by purity ($\epsilon \cdot p$) and relevant vertex detector parameters was quantified. 
The measurement of Higgs$\to b\bar{b}/c\bar{c}/gg$ at the CEPC revealed a correlation between $\rm Tr_{\rm mig}$ and the c-jet tagging $\epsilon \cdot p$. 
By combining these correlations, the relationship between $Tr_{mig}$ and relevant vertex detector parameters can be obtained, shown as the top plot of Fig.~\ref{PNscan}. 
This correlation is formulated in expression~\ref{eq:LCFI}, where $\rm R_{\rm radius}^0$/$\rm R_{\rm resolution}^0$/$\rm R^0_{\rm material}$ represent the default design of CEPC vertex detector and $\rm R_{\rm radius}$/$\rm R_{\rm resolution}$/$\rm R_{\rm material}$ represent the new design. 
The coefficients fitted from the correlations indicate the importance of the corresponding detector parameter on the flavor tagging performance. 
The results obtained from LCFIPlus demonstrate that the flavor tagging performance is more sensitive to the inner radius, followed by the material budget, and lastly the spatial resolution.

\begin{equation}
\label{eq:LCFI}
\begin{split}
\rm Tr_{\rm mig} &= 2.30 + 0.06\cdot \rm log_2\frac{R^0_{material}}{R_{material}} \\
&+ 0.04\cdot log_2\frac{R_{resolution}^0}{R_{resolution}} + 0.10\cdot log_2\frac{R_{radius}^0}{R_{radius}}
\end{split}
\end{equation}

\begin{equation}
\label{eq:PN}
\begin{split}
\rm Tr_{mig} &= 2.64 + 0.03\cdot log_2\frac{R^0_{material}}{R_{material}} \\
&+ 0.02\cdot log_2\frac{R_{resolution}^0}{R_{resolution}} + 0.06\cdot log_2\frac{R_{radius}^0}{R_{radius}} 
\end{split}
\end{equation}

The same analysis was conducted using ParticleNet.
The Z-pole samples are fully simulated based on different vertex detector configurations and fed to ParticleNet to train. 
The results are illustrated in the bottom plot of Fig.~\ref{PNscan} and equation~\ref{eq:PN}.
Compared to LCFIPlus, ParticleNet exhibits a larger $Tr_{mig}$ value (2.64 v.s. 2.30), and its coefficients are roughly 50\% of those of LCFIPlus.
In other words, the ParticleNet has a lower dependence on the geometric parameters.
However, both methods have the same order of impact for three different geometric parameters: both identify the inner radius as the most sensitive to flavor tagging performance and spatial resolution as the least sensitive.

\begin{table}
\caption{ The accuracy of $\nu\nu Hc\bar{c}$ and $V_{cb}$ measurement is assessed under three scenarios: conservative, baseline, and optimal. In the conservative and optimal scenarios, three vertex detector parameters are adjusted to 2 and 0.5 times their values in the baseline design. 
The value of $\rm \frac{LCFIPlus}{ParticleNet}$ reflects the impact of the flavor tagging algorithm on benchmark measurements. }
\label{benchCompare}
\centering
\begin{tabular}{ccccc}
\hline
                                                         &                                                                                         &conservative & baseline   &optimal           \\
\hline
\multirow{3}{*}{$\nu\nu Hc\bar{c}$}  & LCFIPlus                                                                             & 0.071    & 0.057    & 0.047                       \\
                                                         & ParticleNet                                                                         & 0.045   & 0.042     & 0.038                       \\
                                                         & $\rm \frac{LCFIPlus}{ParticleNet}$                                    & 1.58   & 1.36     & 1.26                      \\
\hline
\multirow{3}{*}{$|V_{cb}|$}            & LCFIPlus                                                                   & 0.0241   & 0.0133   & 0.0091                   \\
                                                     & ParticleNet                                                                  & 0.0086  & 0.0076   & 0.0067                   \\
                                                     & $\rm \frac{LCFIPlus}{ParticleNet}$                              & 2.80     & 1.75        &1.36 \\
\hline
\end{tabular}
\end{table}

The influence of geometric modifications on benchmark analyses can be assessed by referring to Fig.~\ref{PNscan} in subsection~\ref{4.1}.
Consider two scenarios: one optimal and the other conservative, where the values of three vertex detector parameters is 0.5/2 times compared to the baseline design.
This adjustment leads to changes in $\rm Tr_{mig}$ from 2.64 to 2.75/2.53 for ParticleNet and from 2.30 to 2.10/2.50 for LCFIPlus, as indicated by the vertical lines in Fig.~\ref{anaPerform}.
The accuracy of $\sigma(ZH)\cdot Br(Z\to \nu\bar{\nu}, H\to c\bar{c})$ and $|V_{cb}|$ measurement under different scenarios using ParticleNet and LCFIPlus is presented in Table~\ref{benchCompare}.
Compared to LCFIPlus, ParticleNet significantly improves the accuracy of benchmark measurements.
In the baseline scenario, the improvement is 36\% and 75\% for $\sigma(ZH)\cdot Br(Z\to \nu\bar{\nu}, H\to c\bar{c})$ and $V_{cb}$ measurement, respectively.
While at the conservative scenario, the improvement can be enhanced to 58\% for $\sigma(ZH)\cdot Br(Z\to \nu\bar{\nu}, H\to c\bar{c})$ and nearly 3 times for $V_{cb}$.

\section{Conclusion}
\label{sec:Con}

Flavor tagging, a methodology employed to discern the origins of jets, holds immense significance in the realm of experimental exploration at the High Energy Frontier.
Jets originating from different quarks or gluons have key differences, represented in the multiplicity of different species of particles, the secondary vertices, the opening angle of jets, etc.
The flavor tagging performance depends on both the flavor tagging algorithm and detector design.
To pursue excellent discovery power and innovative design of the detector, intensive research and design towards the key detector technologies, especially towards the vertex detectors are performed.
Meanwhile, the development of innovative algorithms injects new momentum into this field.

In this paper, we analyze the performance of ParticleNet and LCFIPlus.
The ParticleNet based on GNN has been intensively used at CMS~\cite{PRLQu,PN1,PN2} and FCC-ee~\cite{FCCPN}.
The LCFIPlus is a GBDT-based algorithm that has served as the baseline flavor tagging algorithm for CEPC and multiple future electron-positron Higgs factories.
Using fully simulated hadronic events at a center-of-mass energy of 91.2~GeV at the CEPC baseline detector, we quantify the performance of both algorithms.
We use a 3-dimensional migration matrix to describe the flavor tagging performance (representing the identification efficiency and misidentification rate), and the trace of the migration matrix is used as the key parameter to characterize flavor tagging.

At the CEPC baseline detector geometry, we observe that ParticleNet is significantly superior to LCFIPlus.
At the inclusive hadronic Z pole sample, the trace of ParticleNet is larger than LCFIPlus by more than 14\%.
Consequently, the relative statistical accuracy of $\sigma(ZH)\cdot Br(Z\to \nu\bar{\nu}, H\to c\bar{c})$ and $|V_{cb}|$ measurement via W boson decay is improved by 36\% and 75\%, respectively, when CEPC operates as a Higgs factory at the center-of-mass energy of 240~GeV and integrated luminosity of 5.6~$ab^{-1}$.
Another paper~\cite{Liao:2022ufk} shows that ParticleNet can improve the statistical uncertainty of $R_C$ measurement by 60\% at the CEPC.
The flavor tagging performance, which is described by $Tr_{mig}$, of both ParticleNet and LCFIPlus depends on the polar angle.
Both algorithms exhibit better performance in the barrel and smoothly degrade in the forward region.
We also apply ParticleNet to different vertex detector geometries and observe that the flavor tagging performance is most sensitive to the inner radius, followed by the material budget and the spatial resolution.
The result is consistent with that conducted by LCFIPlus. 
Benchmark performance in two scenarios, conservative and optimal, where the values of three vertex detector parameters are 2 and 0.5 times compared to the baseline design, reveals that ParticleNet can significantly enhance physics performance in the conservative scenario while showing less significant improvement with the aggressive detector design.


\begin{acknowledgements}
We thank the computing center of the Institute of High Energy Physics for providing the computing resources. 
Thanks to Gang Li, Congqiao Li, and Shudong Wang for providing guidance on software.
This project is supported by the Fundamental Research Funds for the Central Universities, Peking University.
This project is also supported by the National Natural Science Foundation of China under grant No. 12042507.
\end{acknowledgements}



\end{document}